\documentclass[11pt,a4paper]{amsart}
\usepackage{amsmath}
\usepackage{amssymb}
\usepackage{amsfonts}
\usepackage[toc,page]{appendix}
\usepackage{hyperref}
\usepackage{color}

\def\={\ =\ }

\newcommand{\be}{\begin{equation}}
\newcommand{\ee}{\end{equation}}
\newcommand{\beq}{\begin{equation}}
\newcommand{\eeq}{\end{equation}}
\newcommand{\bea}{\begin{eqnarray}}
\newcommand{\eea}{\end{eqnarray}}

\def\ba{\begin{eqnarray}}
\def\ea{\end{eqnarray}}

\theoremstyle{plain}

\numberwithin{equation}{section}

\input{tcilatex}

\begin{document}
\title[Polynomial solution of quantum Grassmann matrices]{Polynomial
solution of quantum Grassmann matrices}
\address{Departamento de Matem\'{a}tica, Grupo de F\'{\i}sica Matem\'{a}%
tica, Faculdade de Ci\^{e}ncias, Universidade de Lisboa, Campo Grande, Edif%
\'{\i}cio C6, 1749-016 Lisboa, Portugal.}
\email{tierz@fc.ul.pt}
\author{Miguel Tierz}
\maketitle

\begin{abstract}
We study a model of quantum mechanical fermions with matrix-like index
structure (with indices $N$ and $L$) and quartic interactions, recently
introduced by\ Anninos and Silva. We compute the partition function exactly
with $q$-deformed orthogonal polynomials (Stieltjes-Wigert polynomials), for
different values of $L$ and arbitrary $N$. From the explicit evaluation of
the thermal partition function, the energy levels and degeneracies are
determined. For a given $L$, the number of states of different energy is
quadratic in $N$, which implies an exponential degeneracy of the energy
levels. We also show that at high-temperature we have a Gaussian matrix
model, which implies a symmetry that swaps $N$ and $L$, together with a Wick
rotation of the spectral parameter. In this limit, we also write the
partition function, for generic $L$ and $N,$ in terms of a single
generalized Hermite polynomial.
\end{abstract}

\section{Introduction}

The study of matrix models has been proven to be a very fertile area of
research, with significant impact in fundamental questions of physics. In
particular, its relevance in problems of gauge and string theory, since the
introduction of random matrix theory tools in a gauge theoretic context in 
\cite{Brezin:1977sv}, is so widespread now that it can hardly be summarized
with a few (textbook or review) references, such as \cite{Marino:2005sj}-%
\cite{Pestun:2016zxk}. Indeed, a number of seemingly dissimilar models and
theories often leads to a matrix model, in one way or another. For example,
very recently, in \cite{Anninos:2016klf}, models of quantum mechanical
fermions with matrix-like index structure have been studied, with aims at a
better understanding of the physics of a large number of non-locally
interacting fermionic degrees of freedom. This type of setting, where we
have a bosonic matrix model description of a fermionic (discrete) system has
emerged regularly in recent times, see \cite{Anninos:2014ffa}-\cite%
{berenstein} for example. Likewise, and as pointed out it in \cite%
{Anninos:2016klf}, standard fermionic matrix models were previously studied
in \cite{Makeenko:1993jg}-\cite{Paniak:2000zy}. These are models of the type%
\begin{equation*}
\int_{Gr(N)}d\psi d\overline{\psi }e^{N\mathrm{Tr}V(\psi ,\overline{\psi })},
\end{equation*}%
where $Gr(N)$ denotes a certain Grassmann algebra and $V$ is a polynomial
potential. These fermionic models can be shown to be expressible in terms of
standard, bosonic, random matrix ensembles \cite{Paniak:2000zy}. Likewise,
in the mathematical literature it was exhaustively shown that integrals over
Grassmanians can be formulated in terms of one-matrix model of Hermitian or
unitary type \cite{AvM}. This correspondence between fermionic and bosonic
matrix models holds, in general, not only for large $N$ but for arbitrary $N$
as well \cite{Paniak:2000zy,AvM}.

Interestingly, the fermionic matrix model in \cite{Anninos:2016klf}, which
is to be presented and succinctly described below, admits also a description
in terms of a (bosonic) matrix model, which, we show in this work, can be
analyzed using standard random matrix theory tools. The matrix model turns
out to be of the $q$-deformed type of matrix models which appears in
Chern-Simons theory \cite{Tierz,Tierz:2016zcn}. As we shall see, a
consequence is that the partition function, in the simplest case, can be
analytically characterized, for arbitrary $N$, in terms of a $q$-orthogonal
polynomial, the Stieltjes-Wigert polynomial \cite{Tierz,Tierz:2016zcn},
which can be understood as the polynomial part of a $q$-oscillator
wavefunction. These polynomials are also central in\ Chern-Simons theory but
they are used, as we explain below, in a different way in the
characterization of the fermionic matrix model of \cite{Anninos:2016klf}.

We exploit this analytical characterization to compute a large number of
partition functions for specific values of $N$ and $L$. From these
expressions, the spectrum including the degeneracies of the energy levels
can be immediately read off. The determinants and polynomials are easily
implemented with Mathematica, for example, and a large amount of spectral
data is generated with little computational effort. For example, it is
immediate to study a case such as $L=5$ and $N=10$, whose Hilbert space is
of dimension $2^{50}$. Since the number of states of different energy is
only quadratic in $N$, the full spectra can be written down, even for such a
case. To further illustrate this, a Mathematica file is included.

We explain the most salient features of the spectra, giving some examples
below, while leaving a few more (longer) examples in the Mathematica file,
together with the tools to explore further specific cases. One of the main
aspects of the spectra obtained is the fact that, since the number of states
is exponential in the indices $L$ and $N$, there is an exponential
degeneracy of the energy levels. This is reminiscent of the level structure
of the dual resonance models, where the Hagedorn spectra was found precisely
through an exponential degeneracy of the states \cite{Fubini:1969qb}. This
exponential number of degenerate states occur for levels of mid-energy
value, whereas the ground state and first excited states have a smaller,
polynomial in the indices, degeneracy. The highest energy levels are either
non degenerate or have a very low degeneracy. Mathematically, this is due to
the appearance of $q$-Binomial coefficients and to their unimodality, in the
analytical expressions of the partition functions.

In Section 5, we explain the relationship between the matrix model and its
analytical solution, presented in the previous Sections, with a vector
model, also introduced in \cite{Anninos:2016klf}. For this, we also show the
existence of a useful dual determinant solution to the matrix model.

Finally, a considerable simplification of the generic $(N,L)$ solution
emerges in the high-temperature limit of the model, since this limit reduces
the matrix model to a Gaussian matrix model and the SW polynomials reduce to
Hermite polynomials. At the end of the paper, we analyze this regime in
terms of Hermite polynomials and their generalizations, giving a result for $%
L\times N$ generic in terms of a single polynomial and pointing also out
another duality between $L$ and $N$. We also attempt to explore, in the
Appendix, how the large $N$ behavior with $q$ fixed can be analyzed and
related to results involving Rogers-Ramanujan identities \cite{BP1}-\cite%
{Andrews}.

\section{The model}

Let us now focus on the work \cite{Anninos:2016klf} by Anninos and Silva,
which studies a fermionic matrix model consisting of $NL$ complex fermions $%
\{\psi ^{iA},\bar{\psi}^{Ai}\}$ with $i=1,\ldots ,N$ and $A=1,\ldots ,L$
interacting via a quartic interaction. The indices $i$ and $A$ transform in
the bifundamental of a $U(N)\times U(L)$ symmetry. The thermal partition
function studied in \cite{Anninos:2016klf} is given then by%
\begin{equation}
Z[\beta ]=\int \mathcal{D}\psi ^{iA}\mathcal{D}\bar{\psi}^{iA}e^{-\oint
d\tau \left( \bar{\psi}^{Ai}\dot{\psi}^{iA}-\frac{1}{4L\gamma }\bar{\psi}%
^{Ai}\psi ^{iB}\bar{\psi}^{Bj}{\psi }^{jA}\right) }~,  \label{ZB}
\end{equation}%
where the fermions are anti-periodic in Euclidean time. The $\gamma $ is a
real positive parameter but can be taken to be complex \cite{Anninos:2016klf}%
. The Hilbert space of the theory consists of $N\times L$ fermionic
operators which satisfy $\{\bar{\psi}^{Ai},\psi ^{Bj}\}=\delta ^{ij}\delta
^{AB}$ and the dimension of the space is $2^{N\times L}$. The interaction
Hamiltonian is a quartic term, in which the indices are traced%
\begin{equation*}
H=-(4L\gamma )^{-1}\sum_{A,B,i,j}\bar{\psi}^{Ai}\psi ^{iB}\bar{\psi}%
^{Bj}\psi ^{jA},
\end{equation*}%
together with quadratic normal ordering terms. The analysis of the bosonic
path integrals in \cite{Anninos:2016klf} begins with a Hubbard-Stratonovich
transformation, which introduces a bosonic variable $\lambda (\tau ),$ a $%
N\times N$ \ Hermitian matrix $M_{ij}(\tau )$ transforming in the adjoint of
the $U(N)$. We do not reproduce the details of their derivation here, which
can be found in detail in their Section 4.1 (and in previous Sections there
for their simpler vector model). We focus, as a starting point, on the
result they obtain, which shows that (\ref{ZB}) admits a simple and very
concrete one-matrix model representation%
\begin{equation}
\tilde{Z}=\mathcal{Q}\,\int \prod_{i=1}^{N}d\mu _{i}\prod_{i<j}\sinh
^{2}\left( \frac{\mu _{i}-\mu _{j}}{2}\right) \prod_{i=1}^{N}\,\cosh ^{L}%
\frac{\mu _{i}}{2}\,e^{-L\tilde{\gamma}\mu _{i}^{2}}\,,  \label{sinhZ}
\end{equation}%
with normalization constant%
\begin{equation}
\mathcal{Q}^{-1}=2^{-L}\int \prod_{i=1}^{N}d\mu _{i}\prod_{i<j}\sinh
^{2}\left( \frac{\mu _{i}-\mu _{j}}{2}\right) \prod_{i=1}^{N}e^{-L\tilde{%
\gamma}\mu _{i}^{2}}~.  \label{norm}
\end{equation}%
The parameter satisfies $\tilde{\gamma}>0$. The model derived in \cite%
{Anninos:2016klf} has actually a trigonometric Vandermonde term in the
matrix model, rather than the hyperbolic one in (\ref{sinhZ}), but a change
of variables $\mu _{i}\rightarrow i\mu _{i}$ is used in \cite%
{Anninos:2016klf} to obtain (\ref{sinhZ}). Note that this simple
transformation brings the contour of integration from the real to the
imaginary line. The required rotation of the integration contour back to the
real line was studied, for pure Chern-Simons theory on $S^{3}$, which
corresponds to $L=0$ above, in \cite{Romo:2011qp}. See the Appendix in \cite%
{Romo:2011qp} for explicit evaluations of the partition functions. For $L>0,$
one does not expect additional difficulties, in contrast to some
Chern-Simons-matter matrix models, characterized by (\ref{sinhZ}) with $L\in 
\mathbb{Z}
^{-}$. This implies that there are poles in the integrand, and then rotation
of contours becomes trickier. In that case, the hyperbolic and trigonometric
models will in principle differ \cite{Russo:2014bda}. In this paper, we
follow \cite{Anninos:2016klf} and study (\ref{sinhZ}).

We show here that this model is exactly solvable for $N$ finite. In
particular, the tools used in \cite{Tierz:2016zcn} for Chern-Simons-matter
(CSM) matrix models directly apply here, and they do so in a much simpler
fashion, since the term $\cosh ^{L}\left( \mu _{i}/2\right) $ can be
understood as a characteristic polynomial insertion in the matrix model (\ref%
{norm}), which is a Stieltjes-Wigert ensemble \cite{Tierz}. This insertion,
in contrast to the ones in \cite{Tierz:2016zcn}, is in the numerator and
hence the solution is directly in terms of Stieltjes-Wigert polynomials, and
not their Cauchy transform. As a matter of fact, after a simple rewriting,
some of the relevant computations are contained in much earlier work \cite%
{Dolivet:2006ii}, since to study averages of Schur polynomials in the
ensemble (\ref{norm}), we computed first the averages of correlation
functions of characteristic polynomials in the ensemble (\ref{norm}). As we
shall see, (\ref{sinhZ}) can be understood as a\ coincident limit of such
correlators.

\section{Exact solution}

Let us rewrite (\ref{sinhZ}) to make the characteristic polynomial insertion
more manifest. First, it is immediate that%
\begin{equation}
\tilde{Z}=\mathcal{Q}e^{-\frac{3NL}{16\tilde{\gamma}}}2^{-NL}\int
\prod_{i=1}^{N}d\mu _{i}e^{-L\tilde{\gamma}\sum_{i=1}^{N}\mu
_{i}^{2}}\prod_{i=1}^{N}\,\left( e^{\frac{1}{4\tilde{\gamma}}}+e^{\mu
_{i}}\right) ^{L}\prod_{i<j}\sinh ^{2}\left( \frac{\mu _{i}-\mu _{j}}{2}%
\right) \,\,,  \label{reZ}
\end{equation}%
and the spectral parameter will be identified after bringing the expression
in standard random matrix form. Indeed, using \cite[Eq 2.14]{Tierz}, we have
that%
\begin{equation*}
\tilde{Z}=\mathcal{Q}\widehat{A}\int \prod_{i=1}^{N}dx_{i}e^{-L\tilde{\gamma}%
\sum_{i=1}^{N}\log ^{2}x_{i}}\prod_{i=1}^{N}\,\left( \lambda +x_{i}\right)
^{L}\prod_{i<j}\left( x_{i}-x_{j}\right) ^{2},
\end{equation*}%
where%
\begin{equation}
\widehat{A}=2^{N(N-1)-NL}\exp \left( -\frac{N^{3}}{4L\tilde{\gamma}}-\frac{%
3NL}{16\tilde{\gamma}}-\frac{N^{2}}{2\tilde{\gamma}}\right) ,
\label{constant}
\end{equation}%
and the spectral parameter is 
\begin{equation*}
\lambda =-\exp \left( 1/4\tilde{\gamma}+N/(2L\tilde{\gamma})\right) .
\end{equation*}%
The determinantal formulas for moments and correlation functions of
characteristic polynomials in random matrix ensembles \cite{BZ}, implies
that $\tilde{Z}$, including the normalization factor $\mathcal{Q}$ and for
generic positive integer $L$ and $N$, is given by an $L\times L$ Wronskian
of Stieltjes-Wigert polynomials \cite{Tierz:2016zcn}. That is\footnote{%
The characteristic polynomial insertion is typically $\det (\lambda -X)$ 
\cite{BZ,Tierz:2016zcn}, where $X$ denotes the matrix. Hence, the extra
factor $(-1)^{LN}$, and we have also identified $\lambda $ accordingly, with
a minus sign. The factor $2^{L}$ appears because it is an additional factor
in $Q$ (\ref{norm}).}:%
\begin{equation}
\widehat{Z}=\frac{\tilde{Z}}{2^{L}\widehat{A}}=\frac{(-1)^{LN}}{%
\prod\limits_{j=0}^{L-1}j!}%
\begin{vmatrix}
S_{N}\left( \lambda \right) & S_{N+1}\left( \lambda \right) & ... & 
S_{N+L-1}\left( \lambda \right) \\ 
S_{N}^{\prime }\left( \lambda \right) & S_{N+1}^{\prime }\left( \lambda
\right) & ... & S_{N+L-1}^{\prime }\left( \lambda \right) \\ 
&  &  &  \\ 
S_{N}^{(L-1)}\left( \lambda \right) & S_{N+1}^{(L-1)}\left( \lambda \right)
&  & S_{N+L-1}^{(L-1)}\left( \lambda \right)%
\end{vmatrix}%
,  \label{W}
\end{equation}%
where $S_{N}\left( \lambda \right) $ denotes the \emph{monic}
Stieltjes-Wigert polynomial \cite{Tierz,Dolivet:2006ii,Tierz:2016zcn} of
degree $N$ and $S_{N}^{(k)}\left( \lambda \right) $ its $k$-th derivative.
The explicit expression for the orthonormal SW polynomials \cite%
{Dolivet:2006ii,Tierz:2016zcn} is%
\begin{equation}
P_{n}(x;q)=\frac{(-1)^{n}q^{n/2+1/4}}{\sqrt{(q;q)_{n}}}\sum_{j=0}^{n}\QATOPD[
] {n}{j}_{q}(-1)^{j}q^{j^{2}+j/2}x^{j},\quad n=0,1,\ldots  \label{SW}
\end{equation}%
These are orthonormal with regards to the weight function%
\begin{equation}
\omega _{\mathrm{SW}}\left( x\right) =\frac{1}{\sqrt{\pi }}k\mathrm{e}%
^{-k^{2}\log ^{2}x},  \label{LN1}
\end{equation}%
with $q=e^{-1/2k^{2}}$. We used the standard notation for the $q$-Pochhammer
symbol%
\begin{equation*}
(q;q)_{0}=1,\quad (q;q)_{n}=\prod_{j=1}^{n}(1-q^{j}),\quad n=1,2,\ldots
\end{equation*}%
In our setting, the $q$-parameter is then, $q=\exp (-1/(2\tilde{\gamma}L))$.
Notice that the spectral parameter of the characteristic polynomial, in
terms of the $q$-parameter, is simply $\lambda =-q^{-L/2-N}$. Let us now
explicitly study specific cases of the general formula (\ref{W}). For this,
we need the monic version of (\ref{SW}), which are given by \cite%
{Tierz:2016zcn}%
\begin{equation}
S_{n}\left( x\right) =(-1)^{n}q^{-n^{2}-n/2}\sum_{j=0}^{n}\QATOPD[ ] {n}{j}%
_{q}(-1)^{j}q^{j^{2}+j/2}x^{j}.  \label{S}
\end{equation}

\subsection{$L=1$ case: $\protect\widehat{Z}_{1\times N}$ is a SW polynomial}

The case of generic $N$ and $L=1$ is especially simple, since the partition
function is directly given by a single Stieltjes-Wigert polynomial, as can
be seen from (\ref{W}). Alternatively, this case also follows directly from
realizing that we have an expression of the form%
\begin{equation}
P_{n}(z)=\frac{1}{\tilde{Z}_{n}}\int_{-\infty }^{\infty }\ldots
\int_{-\infty }^{\infty }\left( \prod_{j=1}^{n}(z-x_{j})\right)
\prod_{i<j}\left( x_{i}-x_{j}\right) ^{2}\ \prod_{j=1}^{n}\omega
(x_{j})dx_{j}\text{ }.  \label{averagecharpol:intro}
\end{equation}%
It then follows from a classical calculation of Heine, that $P_{n}$ can be
characterized as the $n$th $\mathit{monic}$ orthogonal polynomial with
respect to the weight function $\omega (x)=e^{-V(x)}$ \cite%
{HRMT,Tierz:2016zcn}. Therefore, the partition function in this case is
directly the monic version of (\ref{SW}), specialized at the spectral point $%
x=\lambda =-q^{-1/2-N}$%
\begin{equation}
\widehat{Z}_{1\times N}=\pi _{N}\left( \lambda \right)
=q^{-N^{2}-N/2}\sum_{k=0}^{N}\QATOPD[ ] {N}{k}_{q}q^{k^{2}-kN}.  \label{Z1N}
\end{equation}%
Notice that there is a symmetry, not only of the $q$-Binomial coefficients
but also of the accompanying term $q^{k^{2}-kN}$, which essentially reduces
the number of different terms by half\footnote{%
Namely, it holds that $\QATOPD[ ] {N}{k}_{q}=\QATOPD[ ] {N}{N-k}_{q}$ and,
at the same time, $q^{k^{2}-kN}=q^{(N-k)^{2}-(N-k)N}$. Thus, instead of $N+1$
different terms in the sum there are $\left( N+1\right) /2$ for $N$ odd and $%
N/2+1$ for $N$ even.}. Therefore:%
\begin{eqnarray*}
\widehat{Z}_{1\times N} &=&2q^{-N^{2}-N/2}(1+\sum_{k=1}^{\left\lfloor \frac{N%
}{2}\right\rfloor }\QATOPD[ ] {N}{k}_{q}q^{k^{2}-kN})\text{ for }N\text{ odd}
\\
\widehat{Z}_{1\times N} &=&q^{-N^{2}-N/2}(2+2\sum_{k=1}^{\left\lfloor \frac{%
N-1}{2}\right\rfloor }\QATOPD[ ] {N}{k}_{q}q^{k^{2}-kN}+\QATOPD[ ] {N}{N/2}%
_{q}q^{-N^{2}/4})\text{ \ for }N\text{ even,}
\end{eqnarray*}%
where $\left\lfloor \cdot \right\rfloor $ denotes the floor function. The
analytical evaluation of the matrix model was not carried out in \cite%
{Anninos:2016klf} (three specific $L\times N$ cases were given instead: $%
1\times 2$, $2\times 2$, and $2\times 3$) and is the main result in this
paper. The canonical partition function of the model is%
\begin{equation*}
\widehat{Z}_{L\times N}=\mathrm{Tr}\left( e^{-\beta H_{L\times N}}\right) ,
\end{equation*}%
and from its explicit evaluation, the energy levels and degeneracies can be
read off. This is also done in \cite{Anninos:2016klf} for their vector model
and for a few instances of the matrix model. Notice that the $q$-Binomial
coefficients are (Gaussian) polynomials in $q$ and the specialization of the
SW polynomial will always give polynomials in $q$ and/or $q^{1/2}$.

We give below all the solutions in terms of the $q$-parameter. The
identification with the thermodynamical and the model parameters is
automatic by considering that%
\begin{equation*}
q=\exp \left( -\frac{\beta }{2\tilde{\gamma}L}\right) ,
\end{equation*}%
where we have restored, as done in \cite{Anninos:2016klf}, the temperature
dependence, by recovering $\beta $. Let us proceed first to examine a few
particular cases from the formulas above, which also allows to check out the
three specific particular cases given in \cite{Anninos:2016klf}. To simplify
the presentation and the discussion, the expressions given will be for $%
\widehat{Z}$ and will not be retrieving constantly the prefactors above. We
also will factorize the expressions such that the ground-state is of zero
energy. We have that%
\begin{eqnarray}
\widehat{Z}_{1\times 2} &=&S_{2}\left( -q^{-5/2}\right) =q^{-5}(3+q^{-1}),
\label{ZL1} \\
\widehat{Z}_{1\times 3} &=&-S_{3}\left( -q^{-7/2}\right)
=2q^{-21/2}(2+q^{-1}+q^{-2}),  \notag \\
\widehat{Z}_{1\times 4} &=&S_{4}\left( -q^{-9/2}\right)
=q^{-18}(5+3q^{-1}+4q^{-2}+3q^{-3}+q^{-4}),  \notag \\
\widehat{Z}_{1\times 5} &=&-S_{5}\left( -q^{-11/2}\right)
=2q^{-67/2}(1+q+3q^{2}+3q^{3}+3q^{4}+2q^{5}+3q^{6}).  \notag
\end{eqnarray}%
Notice that the counting of states, as in \cite{Anninos:2016klf}, taking
into account the degeneracies, matches exactly the dimension of the Hilbert
space, which is $2^{NL}$. It is immediate to see that, in the examples
above, we indeed have $4$, $8$, $16$ and $32$ total states, counting the
degeneracies. This has been checked in all of the cases analyzed, for a
large number of different $L$ and $N$ values, with the computer, up to very
large Hilbert spaces, such as spaces of size $2^{50}$, and it is
straightforward to go beyond that (see Mathematica file). Interestingly, the
number of states of $\mathit{different}$ energy is given, in this $L=1$
case, by%
\begin{eqnarray*}
N^{2}+N-1\text{ for }N &=&1,3,5,7,... \\
N^{2}+N\text{ for }N &=&2,4,6,8,...\text{ }
\end{eqnarray*}%
In general, we will have two expressions for $N$ odd or even in the case
when $L$ is odd, and a single expression for $L$ even cases. The important
aspect, as we show explicitly below, is that, regardless of $L$, the number
of states with different energy levels is always polynomial (a quadratic
polynomial, in all the cases studied) in $N$. Because, the total number of
energy states of the system is $2^{NL}$ and therefore grows exponentially in 
$N$ for a given $L$, it means that the total number of states is encoded in
a exponential degeneracy of the available states. For finite $L$ in general,
and $L=1$ in particular, the degeneracies concentrate on, roughly, the
mid-values of the spectra. In particular, the degeneracy of the ground state
and first excited states of this particular $L=1$ case are given by%
\begin{equation*}
\deg \left( E_{0}\right) =N+1\text{ and }\deg \left( E_{1}\right) =N-1.
\end{equation*}

\subsubsection{Product (Fermionic) form and free energy}

It follows from the $q$-Binomial theorem that%
\begin{equation*}
\left( xq;q\right) _{N}=\sum\limits_{k=0}^{N}\QATOPD[ ] {N}{k}%
_{q}(-x)^{k}q^{k(k+1)/2},
\end{equation*}%
which is also a particular case of the finite version of the product formula
for theta functions, which physically is well-known to be interpreted as
bosonization. It follows that\footnote{%
Notice that the $(-1)^{N}$ prefactor of the SW polynomials always cancels
with the $(-1)^{LN}$ in (\ref{W}).}%
\begin{eqnarray}
\widehat{Z}_{1\times N} &=&q^{-N^{2}-N/2}\left( q^{\frac{1}{2}-N};q\right)
_{N}  \notag \\
&=&q^{-N^{2}-N/2}\prod_{j=0}^{N}(1-q^{j-(N-1/2)}).  \label{produ}
\end{eqnarray}%
An infinite series expansion of the free energy also follows from this
expression by taking the log and Taylor expanding, obtaining that%
\begin{equation*}
F_{N}=-\left( N^{2}+N/2\right) \log q-\sum_{k=1}^{\infty }\frac{%
q^{-Nk/2}\left( 1-q^{(N+1)k}\right) }{k\left( 1-q^{k}\right) }
\end{equation*}%
There also exists well-known bounds on the $q$-Pochhammer symbol \cite{qbin}%
, which implies the following bounds for our partition function 
\begin{equation*}
\widehat{Z}_{1\times N}\leq q^{-N^{2}-N/2}\exp \left( \frac{\mathrm{Li}%
_{2}\left( q^{-N-1/2}\right) -\mathrm{Li}_{2}\left( q^{-1/2}\right) }{\log q}%
\right) ,
\end{equation*}%
and%
\begin{equation*}
\widehat{Z}_{1\times N}\geq q^{-N^{2}-N/2}\left( 1-q^{-N-1/2}\right) \exp
\left( \frac{\mathrm{Li}_{2}\left( q^{-N-1/2}\right) -\mathrm{Li}_{2}\left(
q^{-3/2}\right) }{\log q}\right) ,
\end{equation*}%
and, likewise, the free energy is bounded by dilogarithms. For example, the
upper bound reads, writing $\log q=-\beta /(2\tilde{\gamma}L)$%
\begin{equation*}
F_{1\times N}\leq \frac{\beta }{2\tilde{\gamma}}\left( N^{2}+N/2\right) -%
\frac{2\tilde{\gamma}}{\beta }\mathrm{Li}_{2}\left( e^{\frac{\beta \left(
N+1/2\right) }{2\tilde{\gamma}}}\right) -\mathrm{Li}_{2}\left( e^{\frac{%
\beta }{4\tilde{\gamma}}}\right) .
\end{equation*}

\section{$L=2,3,4$ and $5$ and large $N$}

Let us discuss now, using the determinantal formula and the orthogonal
polynomials a few other cases, for larger values of $L$. At the practical
level this can be done very quickly and efficiently with Mathematica.

\subsection{$L=2$}

In this case, the partition function is given by a $2\times 2$ determinant
(Wronskian). Hence the partition function is%
\begin{equation}
\widehat{Z}_{2\times N}=S_{N}\left( \lambda \right) S_{N+1}^{\prime }\left(
\lambda \right) -S_{N}^{\prime }\left( \lambda \right) S_{N+1}\left( \lambda
\right) ,  \label{L2diag}
\end{equation}%
where the spectral parameter is now $\lambda =-q^{-1-N}.$ This analytical
expression is also the polynomial part of the diagonal of the two-point
kernel of the SW ensemble. In other words, the polynomial part of the
(oscillatory) density of states of the matrix model \cite{deHaro:2005rz},
but evaluated specifically at the values determined by the spectral
parameter $\lambda $. Using the explicit expression (\ref{S}) and its
derivative one can easily write down a formal double-sum polynomial
expression for (\ref{L2diag}) but we rather focus instead on specific and
concrete evaluations of the partition function and the ensuing spectra and
degeneracies.

Two explicit results are given in \cite{Anninos:2016klf}: the $2\times 2$
and $2\times 3$ cases. We check those and generate many more with the
polynomials. For example:%
\begin{eqnarray*}
\widehat{Z}_{2\times 2} &=&q^{-11}\left( 3+\frac{4}{q^{1/2}}+\frac{3}{q}+%
\frac{4}{q^{3/2}}+\frac{1}{q^{2}}+\frac{1}{q^{3}}\right) , \\
\widehat{Z}_{2\times 3} &=&q^{-45/2}\left( 4+\frac{6}{q^{1/2}}+\frac{6}{q}+%
\frac{10}{q^{3/2}}+\frac{8}{q^{2}}+\frac{8}{q^{5/2}}+\frac{8}{q^{3}}+\frac{4%
}{q^{7/2}}+\frac{4}{q^{4}}+\frac{2}{q^{9/2}}+\frac{2}{q^{5}}+\frac{2}{%
q^{11/2}}\right) , \\
\widehat{Z}_{2\times 4} &=&q^{-38}\left( 5+\frac{8}{q^{1/2}}+\frac{9}{q}+%
\frac{16}{q^{3/2}}+\frac{17}{q^{2}}+\frac{20}{q^{5/2}}+\frac{24}{q^{3}}+%
\frac{24}{q^{7/2}}+\frac{22}{q^{4}}+\frac{20}{q^{9/2}}+\frac{22}{q^{5}}+%
\frac{16}{q^{11/2}}\right) \\
&&+q^{-38}\left( \frac{14}{q^{6}}+\frac{12}{q^{13/2}}+\frac{8}{q^{7}}+\frac{8%
}{q^{15/2}}+\frac{5}{q^{8}}+\frac{4}{q^{17/2}}+\frac{1}{q^{9}}+\frac{1}{%
q^{10}}\right) .
\end{eqnarray*}%
Notice that, with this factorization, the first two cases above coincide
exactly with the ones given in \cite{Anninos:2016klf}. The number of
different energy levels in this case is simply given by $N(N+1)$ for both $N$
even and odd. The degeneracies of the ground state and the first excited
states can be found explicitly%
\begin{eqnarray*}
\deg (E_{0}) &=&N+1,\text{ }\deg (E_{1})=2N,\text{ }\deg (E_{2})=3(N-1),%
\text{ } \\
\deg (E_{3}) &=&6(N-2)+4,\text{ }\deg (E_{4})=9(N-3)+8\text{ \ \ for }N\geq
3.
\end{eqnarray*}

\subsection{$L=3$}

The first partition functions are 
\begin{eqnarray*}
\widehat{Z}_{3\times 1} &=&\frac{2}{q^{15/2}}+\frac{6}{q^{13/2}}, \\
\widehat{Z}_{3\times 2} &=&\frac{1}{q^{50}}+\frac{1}{q^{49}}+\frac{7}{q^{48}}%
+\frac{15}{q^{47}}+\frac{20}{q^{46}}+\frac{20}{q^{45}}, \\
\widehat{Z}_{3\times 3} &=&\frac{2}{q^{93/2}}+\frac{8}{q^{91/2}}+\frac{16}{%
q^{89/2}}+\frac{28}{q^{87/2}}+\frac{56}{q^{85/2}}+\frac{82}{q^{83/2}}+\frac{%
96}{q^{81/2}}+\frac{104}{q^{79/2}}+\frac{70}{q^{77/2}}+\frac{50}{q^{75/2}},
\\
\widehat{Z}_{3\times 4} &=&\frac{1}{q^{78}}+\frac{1}{q^{77}}+\frac{9}{q^{76}}%
+\frac{24}{q^{75}}+\frac{52}{q^{74}}+\frac{106}{q^{73}}+\frac{168}{q^{72}}+%
\frac{254}{q^{71}}+\frac{352}{q^{70}}+\frac{458}{q^{69}}+\frac{537}{q^{68}}+%
\frac{555}{q^{67}} \\
&&+\frac{534}{q^{66}}+\frac{443}{q^{65}}+\frac{322}{q^{64}}+\frac{175}{q^{63}%
}+\frac{105}{q^{62}}.
\end{eqnarray*}%
Notice that the spectra is always exactly harmonic, with the $N$ even cases
with energies in the integers and for $N$ odd in the half-integers. This is
typical of the $L$ odd cases (see $L=5$ below) and the $L$ even cases are
characterized by somewhat more intricate and less regular pattern of
energies. The degeneracy of the ground state in this case is given by
four-dimensional pyramid number, which is given explicitly by a quartic
polynomial in $N$. In particular:%
\begin{equation*}
\deg (E_{0}\left( L=3,N\right) )=\frac{1}{12}(2+N)^{2}(3+4N+N^{2}).
\end{equation*}%
Interestingly, the degeneracy of the ground-state is the same for the next
case, corresponding to $L=4$. \ The number of different states ($n=1,2,...$)
in this case is $3n^{2}+2n+1$ for $N=2n,$ and $3n^{2}-n$ for $N=2n-1,$ with $%
n=1,2,3,...$ and therefore also quadratic in $N$.

\subsection{$L=4$}

The first three partition functions are 
\begin{eqnarray*}
\widehat{Z}_{4\times 1} &=&\frac{8}{q^{21/2}}+\frac{2}{q^{12}}+\frac{6}{%
q^{10}}, \\
\widehat{Z}_{4\times 2} &=&\frac{1}{q^{36}}+\frac{1}{q^{35}}+\frac{1}{q^{34}}%
+\frac{8}{q^{67/2}}+\frac{1}{q^{33}}+\frac{8}{q^{65/2}}+\frac{15}{q^{32}}+%
\frac{16}{q^{63/2}}+\frac{27}{q^{31}}+\frac{16}{q^{61/2}}+\frac{40}{q^{57/2}}%
+\frac{40}{q^{59/2}} \\
&&+\frac{27}{q^{30}}+\frac{35}{q^{29}}+\frac{20}{q^{28}}, \\
\widehat{Z}_{4\times 3} &=&\frac{2}{q^{68}}{\small +}\frac{2}{q^{67}}{\small %
+}\frac{8}{q^{\frac{133}{2}}}{\small +}\frac{10}{q^{66}}{\small +}\frac{16}{%
q^{\frac{131}{2}}}{\small +}\frac{16}{q^{65}}{\small +}\frac{32}{q^{\frac{129%
}{2}}}{\small +}\frac{30}{q^{64}}{\small +}\frac{40}{q^{\frac{127}{2}}}%
{\small +}\frac{78}{q^{63}}{\small +}\frac{88}{q^{\frac{125}{2}}}{\small +}%
\frac{126}{q^{62}}{\small +}\frac{128}{q^{\frac{123}{2}}}{\small +}\frac{168%
}{q^{61}}{\small +}\frac{208}{q^{\frac{121}{2}}} \\
&&{\small +}\frac{222}{q^{60}}{\small +}\frac{304}{q^{\frac{119}{2}}}{\small %
+}\frac{278}{q^{59}}{\small +}\frac{296}{q^{\frac{117}{2}}}{\small +}\frac{%
326}{q^{58}}{\small +}\frac{304}{q^{\frac{115}{2}}}{\small +}\frac{328}{%
q^{57}}{\small +}\frac{272}{q^{\frac{113}{2}}}{\small +}\frac{258}{q^{56}}%
{\small +}\frac{232}{q^{\frac{111}{2}}}{\small +}\frac{154}{q^{55}}{\small +}%
\frac{120}{q^{\frac{109}{2}}}{\small +}\frac{50}{q^{54}}{\small .}
\end{eqnarray*}%
Many more cases are included and can be further generated in the Mathematica
file. Exploring those cases, the degeneracy of the ground state and the
first excited state can also be easily found, and it is quartic in $N$ for
all $N$\qquad \qquad 
\begin{eqnarray*}
\deg (E_{0}\left( L=3,N\right) ) &=&\frac{1}{12}(2+N)^{2}(3+4N+N^{2}), \\
\deg (E_{1}\left( L=4,N\right) ) &=&N(N+1)(N+2)(N+3)/3.
\end{eqnarray*}%
The number of states of \textit{different} energies is again quadratic in $N$%
, and for example for $N$ even, is given by $2N^{2}+4N-1$ for $N=2,4,6,...$

\subsection{$L=5$}

Again, as above, we have%
\begin{eqnarray*}
\widehat{Z}_{5\times 1} &=&\frac{2}{q^{35/2}}+\frac{10}{q^{31/2}}+\frac{20}{%
q^{29/2}}, \\
\widehat{Z}_{5\times 2} &=&\frac{1}{q^{50}}+\frac{1}{q^{49}}+\frac{1}{q^{48}}%
+\frac{11}{q^{47}}+\frac{11}{q^{46}}+\frac{33}{q^{45}}+\frac{84}{q^{44}}+%
\frac{84}{q^{43}}+\frac{154}{q^{42}}+\frac{224}{q^{41}}+\frac{245}{q^{40}}+%
\frac{175}{q^{39}}, \\
\widehat{Z}_{5\times 3} &=&{\small q}^{-\frac{147}{2}}\left( \frac{2}{q^{19}}%
+\frac{2}{q^{18}}+\frac{14}{q^{17}}+\frac{44}{q^{16}}+\frac{86}{q^{15}}+%
\frac{136}{q^{14}}+\frac{254}{q^{13}}+\frac{510}{q^{12}}+\frac{768}{q^{11}}+%
\frac{1184}{q^{10}}+\frac{1756}{q^{9}}+\frac{2462}{q^{8}}\right) \\
&&{\small +q}^{-\frac{147}{2}}\left( \frac{3184}{q^{7}}+\frac{3578}{q^{6}}+%
\frac{4200}{q^{5}}+\frac{4284}{q^{4}}+\frac{4032}{q^{3}}+\frac{3332}{q^{2}}+%
\frac{1960}{q}+980\right) .
\end{eqnarray*}%
The degeneracies of the ground state in this case can be found to be given
by 
\begin{equation*}
\deg (E_{0}\left( L=5,N\right) )=\prod\limits_{i=1}^{N}\frac{\left(
i+3\right) \left( i+4\right) \left( i+5\right) }{i\left( i+1\right) \left(
i+2\right) },
\end{equation*}%
which has a number of different combinatorial interpretations. The number of
states of different energy is given, for $N$ even, by dodecahedral numbers%
\begin{equation*}
\left( \frac{N}{2}+1\right) \left( \frac{5N}{2}-4\right) \text{ for }%
N=2,4,6,8,...
\end{equation*}

\section{Relationship with the vector model in \protect\cite{Anninos:2016klf}
and \textit{dual} determinant description}

The discussion of a vector model in \cite{Anninos:2016klf} precedes their
analysis of the matrix model and, in contrast to the latter, the vector
model is fully solved there. The partition function is shown to be given by%
\footnote{%
We switched their notation which uses the label $N$ for the parameter,
instead of $L$, to properly view it as a particular instance of the matrix
model.}%
\begin{equation}
Z\propto \int d\,\mu \cosh ^{L}\frac{\mu }{2}\,e^{-L\gamma \mu ^{2}}
\label{vector}
\end{equation}%
and therefore it is enclosed within the analysis we make of the matrix
model, corresponding to the case $N=1$ and $L$. The SW polynomials and the
determinant implementation with Mathematica includes this vector model then,
by choosing $N=1$ (see below and the Mathematica file). Notice however, that
this is just a Gaussian integral, and after applying the binomial expansion
to the $\cosh ^{L}\mu /2$ insertion in (\ref{vector}), one obtains%
\begin{equation}
Z_{\mathrm{vector}}\left( \beta \right) =\sum_{n=0}^{L}\binom{L}{n}e^{-\beta
\left( L-2n\right) ^{2}/(16L\gamma )}.  \label{Zvect}
\end{equation}%
Note the relationship with the particular instance $\left( L=1,N\right) $ of
the matrix model (\ref{Z1N}), where an analogous expression holds but with $%
q $-Binomial, instead of binomial coefficients. Both expressions are then
related in a $\gamma \rightarrow \infty $ (or $\beta \rightarrow 0$) limit
(which implies $q\rightarrow 1$). In addition, it can be shown, already from
the integral expression (\ref{vector}), that $Z_{\mathrm{vector}}\left(
\beta \right) $ is a Hermite polynomial in that limit. We will study this
high-temperature limit in general at the end of the paper, where a duality
between $L\times N$ and $N\times L$ holds in general. All this is then a
particular case of such a relationship.

A more specific and useful relationship between (\ref{Zvect}) and the
Stieltjes-Wigert polynomial follows from a powerful identity between
determinants. More specifically, between Wronskians and Hankel determinants,
it holds that \cite[Theorem 1]{Leclerc}%
\begin{equation}
\begin{vmatrix}
S_{N}\left( \lambda \right) & S_{N+1}\left( \lambda \right) & ... & 
S_{N+L-1}\left( \lambda \right) \\ 
S_{N}^{\prime }\left( \lambda \right) & S_{N+1}^{\prime }\left( \lambda
\right) & ... & S_{N+L-1}^{\prime }\left( \lambda \right) \\ 
&  &  &  \\ 
S_{N}^{(L-1)}\left( \lambda \right) & S_{N+1}^{(L-1)}\left( \lambda \right)
&  & S_{N+L-1}^{(L-1)}\left( \lambda \right)%
\end{vmatrix}%
=C_{N,L}%
\begin{vmatrix}
q_{L}\left( \lambda \right) & q_{L+1}\left( \lambda \right) & ... & 
q_{L+N-1}\left( \lambda \right) \\ 
q_{L+1}\left( \lambda \right) & q_{L+2}\left( \lambda \right) & ... & 
q_{L+N}\left( \lambda \right) \\ 
&  &  &  \\ 
q_{L+N-1}\left( \lambda \right) & q_{L+N}\left( \lambda \right) &  & 
q_{L+2N-2}\left( \lambda \right)%
\end{vmatrix}%
,  \label{Lec}
\end{equation}%
where $C_{N,L}$ is a constant whose well-known expression we do not need for
our discussion here. This identity holds for any orthogonal polynomial
system. The polynomials on the r.h.s. of (\ref{Lec}) are related to the
original ones, in the l.h.s., by%
\begin{equation}
q_{n}(x)=\sum_{m=0}^{n}\mu _{m}\binom{n}{m}(-x)^{n-m},  \label{q1}
\end{equation}%
where $\mu _{m}$ denotes the moments%
\begin{equation}
\mu _{m}=\int dxx^{m}\omega \left( x\right) ,  \label{mu}
\end{equation}%
of the weight function $\omega \left( x\right) $ of the orthogonal
polynomials in the l.h.s. of (\ref{Lec}). Therefore, combining (\ref{q1})
and (\ref{mu}) it also holds that 
\begin{equation*}
q_{n}(x)=\int dt(t-x)^{n}\omega \left( t\right) .
\end{equation*}%
In our case, the weight function is the SW one (\ref{LN1}) and hence%
\footnote{%
Recall that, to obtain the exact same result, prefactors included, we need
to consider the l.h.s. of (\ref{Lec}) with the\textit{\ monic }%
Stieltjes-Wigert polynomials, together with the normalization constant there.%
}, $\mu _{m}=q^{(m+1)^{2}/2}$. Thus, (\ref{Zvect}) are the polynomials
-after the spectral specialization- that appear in the \textit{dual}
determinant description and come from shifted moments of the log-normal
(Stieltjes-Wigert) measure.

The usefulness of the expression (\ref{Lec}) comes from the fact that the
original determinant was of size $L\times L$, with polynomials of order $N$
as entries, whereas the r.h.s. is a determinant $N\times N$ with polynomials
of order $L$. Thus, cases which are hard with the original formulation (like
large $L$ and low $N$), are now simple in the dual representation. In turn,
the matrix elements in the r.h.s. of (\ref{Lec}) are partition functions of
the vector model and, in particular, we also saw above that the vector model
corresponds to the case $N=1$ and general $L$, a result which is also
manifestly obtained from considering the identity (\ref{Lec}), for precisely
these values. Indeed, for $N=1$, while on the l.h.s. we still have an $L$ by 
$L$ determinant, on the r.h.s. we just have $q_{L}\left( \lambda \right) $,
modulo the constant term.

Notice also that while the number of terms (energy levels) in (\ref{Zvect})
is linear in the parameter $L$, in the matrix model it is quadratic in the
parameter, because of the (Gaussian) polynomial contained in the $q$%
-Binomial. Likewise, for the same reason, whereas the exponential degeneracy
in the vector model is directly encoded in the Binomial coefficients in (\ref%
{Zvect}), the degeneracies are not immediately identified in the matrix
model expressions, such as (\ref{Z1N}). That is \cite{Part}%
\begin{equation*}
\QATOPD[ ] {k+l}{k}_{q}=\sum_{n=0}^{lk}p_{n}\left( l,k\right) q^{n},
\end{equation*}%
where $p_{n}\left( l,k\right) $ is the number of partitions of $n$, which
fit inside a rectangle of size $l\times k$. Hence, 
\begin{equation*}
\widehat{Z}_{1\times N}=q^{-N^{2}-N/2}\sum_{k=0}^{N}\QATOPD[ ] {N}{k}%
_{q}q^{k^{2}-kN}=q^{-N^{2}-N/2}\sum_{k=0}^{N}\sum_{n=0}^{Nk-k^{2}}p_{n}%
\left( N-k,k\right) q^{k^{2}-kN+n},
\end{equation*}%
and therefore the degeneracies in this case will be, in general, combination
of different $p_{n}$.

\section{On the high-temperature limit}

We succinctly show now how some of the formal results above simplify at the
high-temperature limit and how in that limit a certain duality between $L$
and $N$ emerges, due to a fundamental property of characteristic polynomials
of Gaussian random matrix ensembles \cite{MN,Desrosiers:2008tp}. Let us
write the matrix model part in (\ref{reZ}) (that is, excluding the
prefactors), after rescaling its variables%
\begin{equation*}
\overline{Z}=\int
\prod_{i=1}^{N}dy_{i}e^{-\sum_{i=1}^{N}y_{i}^{2}}\prod_{i=1}^{N}\,\left(
\lambda +\exp \left( y_{i}/\sqrt{L\tilde{\gamma}}\right) \right)
^{L}\prod_{i<j}\sinh ^{2}\left( \frac{y_{i}-y_{j}}{2\sqrt{L\tilde{\gamma}}}%
\right) .
\end{equation*}%
Hence, the matrix model, in the high-temperature $\tilde{\gamma}\rightarrow
\infty $ limit, can be approximated as%
\begin{equation}
\overline{Z}_{\tilde{\gamma}\rightarrow \infty }\approx \alpha \int
\prod_{i=1}^{N}dy_{i}e^{-\sum_{i=1}^{N}y_{i}^{2}}\prod_{i=1}^{N}\,\left( 
\overline{\lambda }+y_{i}\right) ^{L}\prod_{i<j}\left( y_{i}-y_{j}\right)
^{2},  \label{high-T}
\end{equation}%
where the prefactor is 
\begin{equation*}
\alpha =2^{N(1-N)}\left( L\tilde{\gamma}\right) ^{N(1-N)/2-LN/2},
\end{equation*}%
and the modified spectral parameter 
\begin{equation*}
\overline{\lambda }=\left( \lambda +1\right) \sqrt{L\tilde{\gamma}}\simeq
\left( 2+1/\left( 4\tilde{\gamma}\right) \right) \sqrt{L\tilde{\gamma}}%
\simeq 2\sqrt{L\tilde{\gamma}}.
\end{equation*}%
The same formulas above hold, but now with Hermite polynomials, which
precisely emerge as the \textit{semiclassical} $q\rightarrow 1$ limit of the
SW polynomials, since these are $q$-Hermite polynomials. In addition, there
is a remarkable explicit evaluation of the Wronskian (\ref{W}), for $N$ and $%
L$ generic, in the Hermite case. Using \cite{DG}%
\begin{equation}
\widehat{Z}_{\tilde{\gamma}\rightarrow \infty }=h_{N}^{(L/2)}(\overline{%
\lambda })  \label{sol}
\end{equation}%
where $h_{N}^{(L/2)}(x)$ are monic generalized Hermite polynomials,
orthogonal with respect to the weight $\left\vert x\right\vert ^{L/2}\exp
(-x^{2}/2)$ and were introduced long ago by Szeg\H{o} \cite{szego} (see \cite%
{DG,Her2} for a more recent account and details). They are expressible in
terms of Laguerre polynomials%
\begin{equation}
h_{2N}^{(L/2)}(x)=(-1)^{L/2}2^{2N}N!L_{N}^{L/2-1/2}(x^{2})\text{ \ and \ }%
h_{2N+1}^{(L/2)}(x)=(-1)^{k}2^{2N+1}N!L_{N}^{L/2+1/2}(x^{2}),  \label{sol2}
\end{equation}%
and hence the solution (\ref{sol}) is completely explicit, by using that%
\begin{equation}
L_{n}^{\alpha }(x)=\sum_{j=0}^{n}\binom{n+\alpha }{n-j}\frac{(-x)^{j}}{j!}.
\label{Lag}
\end{equation}%
In other words, the Wronskian of Hermite polynomials can be expressed as a
single Laguerre polynomial, in contrast to the generic case above discussed,
involving Stieltjes-Wigert polynomials. Notice also that the spectral
parameter, the point where the polynomial is evaluated, is now different. In
particular, in this high-temperature limit the spectral parameter $\overline{%
\lambda }\rightarrow \infty $ which leads to consider the Hermite and
Laguerre polynomials in their natural Plancherel-Rotach asymptotics regime 
\cite{szego}. The details of this will be discussed elsewhere.

The relationship between the vector and matrix models discussed above can be
further analyzed, precisely in terms of Hermite polynomials, giving also a
consistency check of the above. We saw that the vector model corresponds to
the limit of the matrix model when $N=1$. Recall the integral expression for
Hermite polynomials%
\begin{equation*}
H_{n}(t)=\frac{2^{n}}{\sqrt{\pi }}\int_{-\infty }^{\infty }dxe^{-x^{2}}(x\pm
it)^{n},
\end{equation*}%
such integral expression is equivalent (and dual) to the random matrix one
by Heine (\ref{averagecharpol:intro}) and such duality is a particular case
of the duality that is discussed in what follows. But before, notice that 
\begin{equation*}
Z_{\mathrm{vector}}\left( \gamma \right) =C(L,\gamma )\int_{-\infty
}^{\infty }dxe^{-x^{2}}(e^{x/\sqrt{L\gamma }}+e^{1/(4\gamma )})^{L},
\end{equation*}%
therefore, in the large-temperature limit $\gamma \rightarrow \infty $%
\begin{equation*}
Z_{\mathrm{vector}}\left( \gamma \rightarrow \infty \right) =\overline{C}%
(L,\gamma )\int_{-\infty }^{\infty }dxe^{-x^{2}}(2\sqrt{L\gamma }+x)^{L}=%
\overline{C}(L,\gamma )H_{L}(2i\sqrt{L\gamma }).
\end{equation*}%
Thus, this polynomial result and its specialization is consistent with the
limit above. Notice the appearance of the $i$ term in the specialization,
which is due to the fact, explained above in Sect 2., that the matrix model
in \cite{Anninos:2016klf} is Wick-rotated whereas the vector model is not.

Let us also exploit the fact that the matrix model (\ref{high-T}) possesses
a special symmetry. More precisely, a duality, different from the general
one explained above, that makes the average in (\ref{high-T}) (almost)
symmetric in the $L$ and $N$ parameters, meaning that the $L$-th moment of a
characteristic polynomial in a Gaussian Unitary ensemble (GUE) of $N\times N$
is related to the $N$th moment of the characteristic polynomial in a GUE of $%
L\times L$ matrices \cite{MN,Desrosiers:2008tp}. In particular, it holds
that \cite{Desrosiers:2008tp}\footnote{%
A more general result holds, see Proposition 1 and 2 there.}%
\begin{equation*}
\left\langle \det \left( s\pm iX\right) ^{n}\right\rangle _{X\in
GUE_{N}}=e^{-\mathrm{tr}\mathbf{S}^{2}}\left\langle \det \left( Y\right)
^{N}e^{Y\mathbf{S}}\right\rangle _{Y\in GUE_{n}},
\end{equation*}%
where we have kept the notation in terms of averages in GUE ensembles of 
\cite{Desrosiers:2008tp}. Notice that because the matrix $S$ is proportional
to the diagonal matrix in our case, this identity also reads, explicitly, as%
\begin{eqnarray*}
\int \prod_{i=1}^{N}dx_{i}e^{-x_{i}^{2}}\,\left( s+ix_{i}\right)
^{L}\prod_{i<j}\left( x_{i}-x_{j}\right) ^{2} &=&e^{-Ns^{2}}\int
\prod_{i=1}^{L}dy_{i}e^{-y_{i}^{2}+sy_{i}}\,\left( y_{i}\right)
^{N}\prod_{i<j}\left( y_{i}-y_{j}\right) ^{2} \\
&=&e^{-3Ns^{2}/4}\int \prod_{i=1}^{L}dy_{i}e^{-y_{i}^{2}}\,\left( y_{i}+%
\frac{s}{2}\right) ^{N}\prod_{i<j}\left( y_{i}-y_{j}\right) ^{2}.
\end{eqnarray*}%
Notice that there is a multiplication by $i$ in the spectral parameter under
the $L\leftrightarrow N$ exchange%
\begin{equation*}
\overline{Z}_{\tilde{\gamma}\rightarrow \infty }^{N\times L}\left( \lambda
\right) =e^{\frac{i\pi N}{2}+\frac{3N\lambda ^{2}}{4}}\overline{Z}_{\tilde{%
\gamma}\rightarrow \infty }^{L\times N}\left( 2i\lambda \right) ,
\end{equation*}%
and, in addition to such a \textit{Wick-rotation}, there is a numerical
prefactor as well. It would be interesting if this symmetry of the
high-temperature limit can be understood from the original approach and
physics motivation of \cite{Anninos:2016klf}.

\section{Outlook}

There are a number of possible open directions for further work. Let us name
a few. It would be interesting to carry out the spectral analysis with more
detail and precision and compare the exponential degeneracy obtained with
that of other models, such as the dual resonance model. This could also
suggest a string theory interpretation of the quantum Grassmann matrix
models. Of course, being the Hilbert space of the theory not only
finite-dimensional but specifically $2^{NL}$ makes the model interesting
from a quantum information point of view, since it can be interpreted as a
system of $NL$ interacting qubits, whose full spectra we have characterized.
The study of finite-dimensional quantum systems with dimension $p^{n}$ where 
$p$ is a prime and $n$ a natural number enjoy a special status and the case $%
p=2$ in particular has been specifically studied in many works \cite{qinf}.

The results obtained show that the partition functions are mathematically
equivalent to certain amplitudes of non-compact branes in the topological $B$%
-model \cite{Okuyama:2006eb,Hyun:2006dr}. It would be interesting to see if
this relationship can be pursued further to understand, for example, if the
system of bifundamental complex fermions happens to describe the surviving
degrees of freedom of the open string modes connecting a compact brane and
non-compact anti-branes.

Another set of open problems have to do with the polynomial solution. For
example, notice that the fact that the partition function is the SW
polynomial itself (for $1\times N$) or a Wronskian of polynomials, has a
number of implications that we have not actually exploited, because the SW
polynomials posses recurrence relationships, satisfy $q$-difference
equations, and there are also explicit evaluation of generating functions of
the polynomials \cite{JSC}, which could be interpreted as grand-canonical
partition functions.

Another task would be to obtain the full solution of the matrix model for
arbitrary $L$ and $N$. A more general formula was actually given in \cite%
{Dolivet:2006ii} and the exact solution here could follow by simply studying
the diagonal limit of that formula. Another interesting route could be to
study the two-point kernel (whose diagonal limit is (\ref{L2diag}) and
characterizes the $2\times N$ case) and use the fact that it is a
reproducing kernel \cite{HRMT} to construct more general solutions and
relationships between them.

Likewise, it would be interesting if there is any appearance or relevance of
generalized Rogers-Ramanujan identities for the partition function and if
so, if it leads to connections with the character of a Virasoro algebra of a
conformal field theory.

Finally, to study double scaling limits. We have taken large $N$ with $q$
fixed limits, and also commented on other possible scaling limits. Given the
existence of the rich Plancherel-Rotach asymptotics for Hermite and Laguerre
polynomials \cite{szego} and its recent extensions to the $q$-deformed
setting in general, and to SW polynomials in particular \cite{WW1}-\cite%
{IS-Z}, it seems interesting to develop the study of all possible scaling
limits.

\subsection*{Acknowledgements}

Correspondence with Guillermo\ Silva is gratefully acknowledged. This work
is supported by the Fund\~{a}\c{c}ao para a Ci\^{e}ncia e Tecnologia
(program Investigador FCT IF2014), under Contract No. IF/01767/2014.

\newpage \appendix

\section{Large $N$ and Rogers-Ramanujan identities}

We try to guess here how the large $N$ behavior with $q$ fixed can be
analyzed and related to results involving Rogers-Ramanujan identities.
Taking into account that in the large $N$ limit, the $q$-Binomial number
becomes%
\begin{equation*}
\QATOPD[ ] {N}{r}_{q}=\frac{1}{\left( q;q\right) _{r}},
\end{equation*}%
then the limit $N\rightarrow \infty $ gives%
\begin{equation}
\widehat{Z}_{1\times N\rightarrow \infty }=\lim_{N\rightarrow \infty
}q^{-N^{2}-N/2}\sum_{r=0}^{\infty }\frac{q^{r^{2}-Nr}}{\left( q;q\right) _{r}%
}.
\end{equation}%
Notice that we have made the unusual step of only taking the limit
explicitly on the $q$-Binomial coefficient. In that way we can relate our
result with a generalization of the Rogers-Ramanujan identities \cite{BP1}
(based on \cite{GIS})%
\begin{equation}
\sum_{r=0}^{\infty }\frac{q^{r^{2}-Nr}}{\left( q;q\right) _{r}}=\frac{%
e_{N}\left( \frac{1}{q}\right) }{\left( q;q^{5}\right) _{\infty }\left(
q^{4};q^{5}\right) _{\infty }}-\frac{d_{N}\left( \frac{1}{q}\right) }{\left(
q^{2};q^{5}\right) _{\infty }\left( q^{3};q^{5}\right) _{\infty }},
\label{ERR}
\end{equation}%
where the $e_{N}$ and $d_{N}$ are well-known polynomials introduced by Schur 
\cite{BP1,GIS,AKP} which both admit a bosonic and fermionic-like expressions 
\cite{Andrews,BP1}. Let us give instead the recurrence relationship that
they both satisfy%
\begin{equation*}
c_{L}(q)=c_{L-1}(q)+q^{L-1}c_{L-2}(q),\text{ \ \ \ }L\geq 2,
\end{equation*}%
with starting values $d_{0}=0$ and $e_{0}=e_{1}=d_{1}=1$. Schur gave the
values of $e_{\infty }$ and $d_{\infty }$ \cite{AKP}%
\begin{eqnarray*}
d_{\infty }\left( q\right) &=&\frac{1}{\left( q^{2};q^{5}\right) _{\infty
}\left( q^{3};q^{5}\right) _{\infty }}, \\
e_{\infty }\left( q\right) &=&\frac{1}{\left( q;q^{5}\right) _{\infty
}\left( q^{4};q^{5}\right) _{\infty }},
\end{eqnarray*}%
with these, we conjecture that%
\begin{equation*}
\lim_{N\rightarrow \infty }q^{N^{2}+N/2}\widehat{Z}_{1\times
N}=\lim_{N\rightarrow \infty }q^{-\sum_{j=0}^{N}\left( 10j+5\right) }\left( 
\frac{1}{\left( q;q^{5}\right) _{\infty }^{2}\left( q^{4};q^{5}\right)
_{\infty }^{2}}-\frac{1}{\left( q^{2};q^{5}\right) _{\infty }^{2}\left(
q^{3};q^{5}\right) _{\infty }^{2}}\right) ,
\end{equation*}%
which leads to the expression 
\begin{equation*}
\lim_{N\rightarrow \infty }q^{6N^{2}+21N/2+5}\widehat{Z}_{1\times N}=\frac{1%
}{\left( q;q^{5}\right) _{\infty }^{2}\left( q^{4};q^{5}\right) _{\infty
}^{2}}-\frac{1}{\left( q^{2};q^{5}\right) _{\infty }^{2}\left(
q^{3};q^{5}\right) _{\infty }^{2}}.
\end{equation*}%
There exists the finite $N$ version of (\ref{ERR}), but its explicit form
involves a somewhat different $q$-Binomial coefficient to that of (\ref{Z1N}%
). It would be interesting to obtain a finite (extended) Rogers-Ramanujan
identity for the partition function (\ref{Z1N}) at $N$ finite. Likewise, if
we study the large $N$ limit of more general partition functions, such as $%
\widehat{Z}_{2\times N}$, given by (\ref{L2diag}), it seems possible that
multi-dimensional generalizations of the Rogers-Ramanujan identities, such
as the Andrews-Gordon identities \cite{Andrews,BP1} emerge or play a role.
This is an interesting open direction.


\end{document}